\documentclass{appolb}
\usepackage{graphicx,amsmath,amssymb,bbm,mathtools}

\usepackage{url}
\usepackage{breakurl}
\usepackage[breaklinks]{hyperref}

\newcommand{\ZZ}{{\mathbbm{Z}}}

\newtheorem{proposition}{Proposition}
\newtheorem{conjecture}{Conjecture}

\begin{document}

\title{Explorations of ternary cellular automata and ternary density classification problems
}
\author{Henryk Fuk\'s$^1$ and Roman Procyk$^2$
\address{$^1$Department of Mathematics and Statistics, Brock University\\
St. Catharines, Ontario L2S 3A1, Canada\\
$^2$Department of Physics, McGill
  University\\Montreal, Quebec H3A 2T8, Canada
}
}
\maketitle
\begin{abstract}
While  binary  nearest-neighbour  cellar  automata  (CA)  have  been  studied  in  detail  and
from many different angles, the same cannot be said about ternary (three-state) CA rules.
We  present some results of our explorations of a small subset of the vast space of ternary rules, namely rules possessing additive invariants.  We first enumerate rules with four different
additive invariants, and then we investigate if any of them could be used to construct
a two-rule solution of generalized density classification problem (DCP). We show that neither simple nor 
absolute classification is possible with a pair of ternary rules where the first rule is all-conserving
and the second one is reducible to two states.
Similar negative result holds for another version of DCP we propose: symmetric interval-wise DCP.
Finally we show an example of a pair of rules which solve non-symmetric interval-wise DCP for initial
configurations containing at least one zero.
\end{abstract}

\section{Introduction}
In both the theory and applications of cellular automata, a lot of effort has been invested in studying
binary rules. Given the importance of binary logic and binary arithmetic in todays computing, this is of course
quite understandable. 

Nevertheless, the next possible arithmetical and logical systems in terms of the number of allowed states,
namely the ternary arithmetic and ternary logic, enjoyed some (if only limited) popularity in the past.
To give concrete examples,  the wooden calculating machine  \cite{Vass} built by Thomas Fowler in 1840 operated in balanced ternary 
arithmetic, using three digits -1, 0 and 1. In the early 20th century, Polish mathematician Jan {\L}ukasiewicz invented and formalized a three-valued logic \cite{Lukasiewicz},
and many followers developed his ideas in subsequent years. In the  second half of the 20th century, experimental 
computers based on ternary arithmetic were constructed, beginning with the the most famous one, the \emph{Setun} computer developed in 1958 at the
 Moscow State University \cite{setun}. In 1980's, ternary CMOS memory chips (ROM) were built 
 at Queens University in Canada. In the most recent times, the possibility of quantum computing based  on qtrits instead qbits
has been suggested \cite{qtrits}.

All of the above indicates that ternary arithmetic and ternary logic are still interesting areas to explore, and that they
may have some application potential.  This inspired us to probe into the huge space of ternary cellular automata, with
the hope of finding some interesting rules, possibly applicable to solving computational problems such as the density classification problem.

Of course, even if one considers only nearest-neighbour ternary cellular automata, the number of possible rules 
is huge, $3^{3^3} \approx 7.63\cdot 10^{12}$. There is no hope to study all of them systematically, as
it has been done in the case of elementary (binary nearest-neighbour) cellular automata, which form a small set of 256 rules.
For this reason, we have decided to take a closer look at only a small subset of ternary rules, namely those which obey simple 
conservation laws, hoping to find some which would be useful in constructing solutions of computational problems similar
to classical density classification problem (see \cite{Oliveira2014} for review and references).

The paper is organized as follows. First we enumerate ternary nearest-neighbour rules with various additive invariants. Then we
investigate if any of the rules found could be used to solve a generalization of the density classification problem to 3 states.

Let us define some basic concepts first. We will be mostly concerned with the space $S=\{0,1,2\}^{L}$ of finite-length configurations of length $L$ . We will impose periodic boundary conditions on configurations $\mathbf{x} \in S$, so that 
for $\mathbf{x}=(x_0, x_1, \ldots x_{L-1})$, the index $i$ in $x_i$ is to be  always  taken modulo $L$, i.e., 
$i\in \ZZ/L$.
The  \emph{local function} (also called \emph{local transition function} or \emph{local rule}) of a ternary nearest-neighbour cellular 
automaton will be a function  $f:\{0,1,2\}^3 \to \{0,1,2\}$, 
while the corresponding global function $F: S \to S$
will be defined as
\begin{align}
 (F(\mathbf{x}))_i=f(x_{i-1}, x_i, x_{i+1})
\end{align} 
for all $i\in \ZZ/L$.
\section{Conservation laws}
Interesting classes of ternary rules are those with additive invariants.
Let $\Psi(x)$ be some function of $x\in \{0,1,2\}$. If, for any periodic configuration $\mathbf{x}$
of length (period) $L$ we have
$\sum_{i=0}^{L-1} \Psi(x_i) = \sum_{i=0}^{L-1} \Psi( (F(x))_i ) $, then we say that
$F$ conserves $\Psi$. 

According to a well-known theorem of Hattori and Takesue \cite{Hattori91}, $\Psi$ is conserved if and only if for all $x_1,x_2,x_3\in \{0,1,2\}$ we have
\begin{multline} \label{takesue}
 \Psi(f(x_0,x_1,x_2))-\Psi(x_0)= \Psi(f(p,p,x_1)) \\+ \Psi(f(p,x_1,x_2)) -\Psi(f(p,p,x_0)) -\Psi(f(p,x_0,x_1)),
\end{multline}
where $p$ is an arbitrarily chosen element of the alphabet $\{0,1,2\}$. For the sake of convenience, 
we will use $p=0$ in all what follows.

Eq. (\ref{takesue}) can be understood as a sort of discrete form of the continuity equation
$\partial \rho/\partial t=- \partial j /\partial x$ describing transport of some quantity with density $\rho$,
where $j$ is the flux of this quantity. It can be transformed to another, equivalent form, where
on the left hand side one has the local  change of $\Psi$ in one time step, analogous to $\partial \rho/\partial t$, and on the right hand side
one has an expression analogous to the spatial derivative of the flux of $\Psi$ (see \cite{Hattori91}
for further details).

Hattori-Takesue theorem makes it quite easy to check if a given ternary rule $f$ conserves $\Psi$ -- all one needs to do is to 
verify the above condition for all $3^{3}=27$ possible sets of values of $x_1,x_2,x_3\in \{0,1,2\}$.

Examples of some possible choices of $\Psi$ are: 
\begin{itemize}
\item[(a)] $\Psi_0(x)=\frac{1}{2}(1-x)(2-x)$, number of cells in state 0 is conserved;
\item[(b)] $\Psi_1(x)=x(2-x)$, number of cells in state 1 is conserved; 
\item[(c)] $\Psi_2(x)=\frac{1}{2}x(x-1)$, number of cells in state 2 is conserved;
\item[(d)] $\Psi_s(x)=x$, sum of cell values is conserved (so-called \emph{number-conserving rule}).
\end{itemize}
Note that conservation of any two of the above implies conservation of the remaining two. Rules which conserve
all (a)--(d) will be called \emph{all-conserving}.

\section{Enumeration}
We will first enumerate ternary rules  conserving various $\Psi$s. The following 
proposition provides a summary of such enumeration.
\begin{proposition}
  Among nearest-neighbour cellular automata with three states, there is
exactly 
\begin{itemize}
 \item[(a)] $9 \cdot 2^{18} = 2 359 296$ rules conserving  
$\Psi_0$ (the same applies to $\Psi_1$ and $\Psi_2$); 
 \item[(b)] 144 rules conserving $\Psi_s$; 
 \item[(c)] 15 all-conserving rules. 
\end{itemize}
 \end{proposition}

\emph{Proof:}\\
Claim (b) is a known results \cite{paper12}, thus we will prove only (a) and (c).
Let $f:\{0,1,2\}^2 \to \{0,1,2\}$ be a local function of a CA conserving 1's. According to the Hattori-Takesue
theorem, it must satisfy
\begin{multline} \label{htcond}
\Psi_1(f(x_0,x_1,x_2))-\Psi_1(x_0)= \Psi_1(f(0,0,x_1)) \\+ \Psi_1(f(0,x_1,x_2)) -\Psi_1(f(0,0,x_0)) -\Psi_1(f(0,x_0,x_1))
\end{multline}
for all $x_1,x_2,x_3\in \{0,1,2\}$, where $\Psi_1(x)=x(2-x)$. 

When values of arguments of $f$ are restricted to  only 0's and 1's, $f$ must return
outputs  compatible with one of the five number-conserving elementary cellular automata (ECA) rules with Wolfram 
numbers 184, 226, 170, 240 or 204, where compatibility is defined as follows. We say that ternary rule
$f$ is \emph{compatible} with binary rule $g$ if, for 
$x_1,x_2,x_3 \in \{0,1\}$,
\begin{equation}
f(x_1,x_2,x_3) = \begin{cases}
1 & \text{if $g(x_1,x_2,x_3)=1$,}\\
0 \text{\,\,or\,\,} 2 &\text{if $g(x_1,x_2,x_3)=0$.}
\end{cases}
\end{equation}
 In the above, we have ``0 or 2'' because we only want to 
preserve the number of 1's.

We will first enumerate rules which, when their arguments are restricted to 0's and 1's, are compatible with rule 184. 
Let us define, for $x_1,x_2,x_3\in \{0,1,2\}$,
\begin{equation}
 a_{9x_0+3x_1+x_2}=f(x_0,x_1,x_2).
\end{equation}
For rules compatible with rule 184 we must have
\begin{align*}
a_{0} =&f(0,0,0)=0 \text{\,\,or\,\,} 2,\\
a_{1} =&f(0,0,1)=0 \text{\,\,or\,\,} 2,\\
a_{3} =&f(0,1,0)=0 \text{\,\,or\,\,} 2,\\
a_{4} =&f(0,1,1)=1,\\
a_{9} =&f(1,0,0)=1,\\
a_{10}=&f(1,0,1)=1,\\
a_{12}=&f(1,1,0)=0 \text{\,\,or\,\,} 2,\\
a_{13}=&f(1,1,1)=1.
\end{align*}
Note that the above implies $\Psi_1(a_{0})=\Psi_1(a_{1})=\Psi_1(a_{3})=\Psi_1(a_{12})=0$.
The remaining $a_i$ must satisfy 27 conditions obtained from eq. (\ref{htcond}). All of them  are listed below
in two columns, including  those which reduce to identities.
\begin{equation*}
    \begin{aligned}[t]
    0=&0,\\
0=&0,\\
\Psi_1\left( a_{{2}} \right) =&\Psi_1\left( a_{{2}} \right) ,\\
0=&0,\\
1=&1,\\
\Psi_1 \left( a_{{5}} \right) =&\Psi_1\left( a_{{5}} \right) ,\\
\Psi_1 \left( a_{{6}} \right) =&\Psi_1\left( a_{{6}} \right) ,\\
\Psi_1 \left( a_{{7}} \right) =&\Psi_1\left( a_{{7}} \right) ,\\
\Psi_1\left( a_{{8}} \right) =&\Psi_1\left( a_{{8}} \right) ,\\
0=&0,\\
0=&0,\\
\Psi_1\left( a_{{11}} \right) -1=&\Psi_1\left( a_{{2}} \right) ,\\
-1=&-1,\\
0=&0,
    \end{aligned} \qquad \qquad    
    \begin{aligned}[t]
 \Psi_1\left( a_{{14}} \right) -1=&-1+\Psi_1\left( a_{{5}} \right) ,\\
\Psi_1\left( a_{{15}} \right) -1=&\Psi_1\left( a_{{2}} \right) +\Psi_1\left( a_{{6}
} \right) -\Psi_1\left( a_{{5}} \right) ,\\
\Psi_1\left( a_{{16}} \right) -1=&\Psi_1\left( a_{{2}} \right) +\Psi_1\left( a_{{7}
} \right) -\Psi_1\left( a_{{5}} \right) ,\\
\Psi_1\left( a_{{17}} \right) -1=&\Psi_1\left( a_{{2}} \right) +\Psi_1\left( a_{{8}
} \right) -\Psi_1\left( a_{{5}} \right) ,\\
\Psi_1\left( a_{{18}} \right) =&-\Psi_1\left( a_{{2}} \right) -\Psi_1\left( a_{{6}}
 \right) ,\\
\Psi_1\left( a_{{19}} \right) =&-\Psi_1\left( a_{{2}} \right) -\Psi_1\left( a_{{6}}
 \right) ,\\
\Psi_1\left( a_{{20}} \right) =&-\Psi_1\left( a_{{6}} \right) ,\\
\Psi_1\left( a_{{21}} \right) =&-\Psi_1\left( a_{{2}} \right) -\Psi_1\left( a_{{7}}
 \right) ,\\
\Psi_1\left( a_{{22}} \right) =&1-\Psi_1\left( a_{{2}} \right) -\Psi_1\left( a_{{7}
} \right) ,\\
\Psi_1\left( a_{{23}} \right) =&\Psi_1\left( a_{{5}} \right) -\Psi_1\left( a_{{2}}
 \right) -\Psi_1\left( a_{{7}} \right) ,\\
\Psi_1\left( a_{{24}} \right) =&\Psi_1\left( a_{{6}} \right) -\Psi_1\left( a_{{8}}
 \right) ,\\
\Psi_1\left( a_{{25}} \right) =&\Psi_1\left( a_{{7}} \right) -\Psi_1\left( a_{{8}}
 \right) ,\\
\Psi_1\left( a_{{26}} \right) =&0.
    \end{aligned}                             
    \end{equation*}
From the above, if we discard all equations which are identically true and
introduce variables  $b_i=\Psi_1(a_i)$, we obtain
\begin{equation*}
    \begin{aligned}[t]
      b_{ {{11}} } -1=&b_{ {{2}} } ,\\
b_{ {{14}} } -1=&-1+b_{ {{5}} } ,\\
b_{ {{15}} } -1=&b_{ {{2}} } +b_{ {{6}
} } -b_{ {{5}} } ,\\
b_{ {{16}} } -1=&b_{ {{2}} } +b_{ {{7}
} } -b_{ {{5}} } ,\\
b_{ {{17}} } -1=&b_{ {{2}} } +b_{ {{8}
} } -b_{ {{5}} } ,\\
b_{ {{18}} } =&-b_{ {{2}} } -b_{ {{6}}
 } ,\\
b_{ {{19}} } =&-b_{ {{2}} } -b_{ {{6}}
 } ,
    \end{aligned} \qquad \qquad    
    \begin{aligned}[t]
 b_{ {{20}} } =&-b_{ {{6}} } ,\\
b_{ {{21}} } =&-b_{ {{2}} } -b_{ {{7}}
 } ,\\
b_{ {{22}} } =&1-b_{ {{2}} } -b_{ {{7}
} } ,\\
b_{ {{23}} } =&b_{ {{5}} } -b_{ {{2}}
 } -b_{ {{7}} } ,\\
b_{ {{24}} } =&b_{ {{6}} } -b_{ {{8}}
 } ,\\
b_{ {{25}} } =&b_{ {{7}} } -b_{ {{8}}
 } ,\\
b_{ {{26}} } =&0.
    \end{aligned}                             
    \end{equation*}
This is a linear system of 14 equations with 19 unknowns. We can, therefore, solve this 
system for $b_{11}, b_{14}, b_{15}, \ldots b_{26}$ in terms of $b_2,b_5, b_6, b_7, b_8$,
obtaining
 \begin{equation*}
    \begin{aligned}[t]
        b_{{11}}&=1+b_{{2}},\\b_{{14}}&=b_{{5}},\\b_{{15}}&=-b_{{5}}+1+b_{{2
}}+b_{{6}},\\b_{{16}}&=-b_{{5}}+1+b_{{2}}+b_{{7}},\\b_{{17}}&=-b_{{5}}+1+b_{
{2}}+b_{{8}},\\b_{{18}}&=-b_{{2}}-b_{{6}},\\b_{{19}}&=-b_{{2}}-b_{{6}},
    \end{aligned} \qquad \qquad    
    \begin{aligned}[t]
   b_{{
20}}&=-b_{{6}},\\b_{{21}}&=-b_{{2}}-b_{{7}},\\b_{{22}}&=1-b_{{2}}-b_{{7}},\\b_{
{23}}&=b_{{5}}-b_{{2}}-b_{{7}},\\b_{{24}}&=b_{{6}}-b_{{8}},\\b_{{25}}&=b_{{7}
}-b_{{8}},\\b_{{26}}&=0. 
    \end{aligned}                             
    \end{equation*}

Recall that the only allowed values of $b_i$ are 0 or 1, thus we must have 
$b_2=0$. By the same token, we  also need $b_7=0$, $b_8=0$, and $b_6=0$.
This leaves
 \begin{equation*}
    \begin{aligned}[t]
      b_{{11}}&=1,\\b_{{14}}&=b_{{5}},\\b_{{15}}&=-b_{{5}}+1,\\b_{{16}}&=-b_{
{5}}+1,\\b_{{17}}&=-b_{{5}}+1,\\b_{{18}}&=0,\\b_{{19}}&=0,
    \end{aligned} \qquad \qquad    
    \begin{aligned}[t]
    b_{{20}}&=0,\\b_{{21}}&=0
,\\b_{{22}}&=1,\\b_{{23}}&=b_{{5}},\\b_{{24}}&=0,\\b_{{25}}&=0,\\b_{{26}}&=0,
    \end{aligned}                             
    \end{equation*}
meaning that we have only one parameter left, $b_5$. Obviously, $b_5\in\{0,1\}$, thus we obtain two
solutions of our original system. The first corresponds to
\begin{align*}
b_i&=0 \,\,\,\text{for}\,\,\,\, i \in \{2,5,6,7,8,14,18 , 19 , 20 ,  21 ,23, 24 ,  25 , 26 \},\\
b_i&=1 \,\,\,\text{for}\,\,\,\, i \in \{11,15,16,17,22\},
\end{align*}
and the second to
\begin{align*}
b_i&=0 \,\,\,\text{for}\,\,\,\, i \in \{2,  6, 7, 8, 15 , 16 , 17 ,18 , 19 , 20 , 21 , 24 , 25 , 26\},\\
b_i&=1 \,\,\,\text{for}\,\,\,\, i \in \{5, 11, 4, 22 , 23\}.
\end{align*}
In both the above solutions, the number of $b_i$'s with zero values is exactly 14. Each $b_i=0$
admits two corresponding values of $a_i$, $a_i=0$ or $a_i=2$, because we defined
$b_i=\Psi_1(a_i)=a_i(2-a_i)$. Moreover, values of 4 parameters $a_0,a_1,a_2$, and $a_{12}$ also admit two possible
values, 0 or 2. 
 This means that each of these two solutions corresponds to $2^{18}$ 
possible sets $a_0, a_2, \ldots a_{26}$ satisfying 27 Hattori-Takesue conditions. We thus have total
$2^{19}$ CA rules conserving 1's and compatible with rule 184.

For rules compatible with  ECA 226, the calculations are almost identical, yielding also $ 2^{19}$ rules.
For rules compatible with  ECA 170 or 240, by using similar reasoning as above, we obtain in both cases $2^{18}$ rules.
For rules compatible with ECA 204, the total number of rules turns out to be $3 \cdot 2^{18}$, again by a similar reasoning
(omitted here).

The total number of rules conserving 1's is, therefore, $2^{19}+2^{19}+2^{18}+2^{18}+3\cdot 2^{18}=9 \cdot 2^{18}$,
as claimed in (a).

For part (c), we  took the advantage of the fact that every all-conserving rule must also 
be number-conserving. One can, therefore, simply test all 144 number-conserving rules for
conservation of the number of 0's (the numbers of 1's and 2's will then be automatically conserved,
so they do not even need to be checked). 
We carried out his procedure and found that 15 rules shown in Table~\ref{allconserving15} are the only all-conserving ternary rules.
\section{Classification problems}
The following two-rule solution of the density classification problem is well known \cite{paper4}. 
\begin{proposition}[H.F. 1997]
Let $\mathbf{x}$ be a periodic binary configuration of length $L$ and density $\rho=\frac{1}{L}\sum_{i=0}^{L-1} x_i$,
 and let $n=\lfloor (L-2)/2 \rfloor$, $m=\lfloor (L-1)/2 \rfloor$. Then 
\begin{itemize}
 \item $F_{232}^m(F_{184}^{n}(\mathbf{x}))=0^L$ if $\rho<1/2$,
  \item $F_{232}^m(F_{184}^{n}(\mathbf{x}))=1^L$ if $\rho>1/2$,
 \item $F_{232}^m(F_{184}^{n}(\mathbf{x}))=\ldots01010101\ldots$ if $\rho=1/2$.
\end{itemize}
\end{proposition}
In the above, $F_{184}$ and  $F_{232}$ are global functions of elementary cellular automata 184 and 232,
and $0^L$ ($1^L$) denotes a string of length $L$ of all zeros (all ones).

 We say that the pair
of rules (184, 232) classifies densities, or solves the \emph{density classification problem} (DCP), because iterating rule 184 sufficient number of times followed by analogous iteration
of rule 232 produces homogeneous string of all zeros if initially we had more zeros than ones, and
homogeneous string of all ones if we had more ones than zeros at the beginning. It is worth noting that 
the above two-rule solution of DCP has been proposed because single-rule solution of this problem does not exist \cite{LB95, busic2013}.

There are three obvious ways to generalize the density classification problem to 3 states (or more).
\begin{itemize}
 \item \emph{0-majority:} in the final configuration all sites are to be in state 0 if
there are more zeros than other symbols in the initial configuration. If there are more non-zero symbols than zeros in the initial configuration, then in the final state  all sites are to be in state 1.
 \item \emph{simple majority:} in the final configuration  all sites are to be in state  $k$ if symbols $k$ form the  majority
in the initial configuration.
 \item \emph{absolute majority:} in the final configuration  all sites are to be in state  $k$ if symbols $k$ form the   absolute majority
in the initial configuration.
\end{itemize}
Note that all three are compatible with the binary DCP, meaning that  when the initial configuration contains only 0's and 1's,
they reduce to the binary DCP.

Since the binary DCP has no single-rule solution, the same applies to all three generalized problems as well.
But are there any two-rule solutions for these problems? In 1999, H.F. Chau \emph{et al.} constructed 
two-rule solution to $n$-ary  simple majority problem \cite{Chau1999}. Their solution, however,
uses rules of rather large neighbourhood size. For ternary rules it requires that the first rule has neighbourhood radius
of at least 45. Despite the fact that Chau's solution uses some very interesting ideas,  we will not discuss
it here, as we wish to focus on nearest-neighbour rules only. Along the same vein, since we restrict our attention to
ternary rules only, we will not 
be concerned with solutions of density classification problems which require very large number states,
such as, for example, the work of Brice\~{n}o \emph{et al.} \cite{BRICENO201370}.

Since in the known two-rule solution of the binary DCP 
the first rule conserves the number of zeros (and ones), it is reasonable to expect that for ternary rules,
the first rule of the solution should also conserve the relevant quantity. For 0-majority problem, the fist rule
should thus be $\Psi_0$-conserving, and for the simple or absolute majority problem, it should be all-conserving. 
Since we have enumerated various rules with invariants, we can search  among the relevant rules for a possible
candidate for the first rule of the solution.

For the 0-majority problem , we do not even have to search among the $9\cdot 2^{18}$ $\Psi_0$-conserving rules, because the solution
is trivial to construct. 
Let
\begin{equation}
f(x_1,x_2,x_3)=f_{184}(\phi(x_1),\phi(x_2),\phi(x_3)) \text {\,\, for all \,\,} x_1,x_2,x_3 \in \{0,1,2\}, 
\end{equation}
where we define $\phi(0)=0$, $\phi(1)=1$, $\phi(2)=1$ and where  $f_{184}$ is the local rule ECA 184.
The pair  $f$ and ECA 232 solve 0-majority DCP in the same fashion as rules
184 and 232 solve the binary DCP. Obviously, one could also define two other variants of the  0-majority problem
(1-majority and 2-majority), and construct their two-rule solutions in an analogous way.

\section{Simple and absolute  density classification}
For the remaining two versions of DCP, we have not found any pair of ternary rules solving these versions. 
We strongly suspect that such a pair does not exist, although we were able to prove only
a partial non-existence result, to be presented below.

Let us  define, for a given ternary rule $f$,  three functions
\begin{align} \label{binaryproj}
f|_{01}(x_1,x_2,x_3)&=f(x_1,x_2,x_3),  \nonumber \\
f|_{02}(x_1,x_2,x_3)&=f(2x_1,2x_2,2x_3)/2, \nonumber \\
f|_{12}(x_1,x_2,x_3)&=f(x_1+1,x_2+1,x_3+1)-1, 
\end{align}
where $x_1,x_2,x_3 \in \{0,1\}$.  
If any of these functions  returns only values 0 or 1 for all  $x_1,x_2,x_3 \in \{0,1\}$,
we will say that the relevant \emph{binary reduction} of $f$ exists. If all three binary reductions exist, we will call $f$
\emph{reducible to two states}.

\begin{proposition}
 There exists no  pair of nearest-neighbour ternary rules  solving the simple (or absolute) majority density classification task such that the first rule of the pair is all-conserving and the second rule is
reducible to two states.
\end{proposition}

\emph{Proof:} Suppose that there exist such a pair of ternary rules solving the simple majority density classification, and the first rule $f$ in this pair is all-conserving, while the second one is reducible to two states.
Table~\ref{allconserving15} shows Wolfram numbers of binary projections for all 15 all-conserving rules.
One can clearly see that  the first rule, when restricted to binary configurations, behaves as one of the
five rules among 204 (identity), 170 (left shift), 240 (right shift), 184, and 226.
Since identity and shifts do not change the arrangement of symbols, $f$ behaving 
as rule 204, 170 or 240 on on binary configurations would require that the second rule performed the entire task of the density classification on its own ($f$ would do nothing).  We assumed reducibility of the second rule, thus
its reduction to $\{0,1\}$ would have to be a solution  of the binary classification problem.
This, however is impossible -- single-rule solution of the binary classification problem does not exist.

Therefore, when we restrict configurations to 
$\{0,1\}$, the all-conserving rule $f$ must satisfy 
\begin{align}\label{cond1}
\text{$\forall x_1,x_2,x_3 \in \{0,1\}$: } f(x_1,x_2,x_3)&=f_{184}(x_1,x_2,x_3) , \text{\,\,or\,\,} \nonumber \\
 \text{$\forall x_1,x_2,x_3 \in \{0,1\}$: } f(x_1,x_2,x_3)&=f_{226}(x_1,x_2,x_3). 
\end{align}
This is equivalent to saying that 
\begin{equation} \label{c1}
f|_{01}=f_{184} \,\,\,\text{or}\,\,\, f|_{01}=f_{226}.
\end{equation}
Exactly the same reasoning applies in the case of reductions to $\{0,1\}$,
yielding the requirement 
\begin{equation} \label{c2}
f|_{02}=f_{184} \,\,\,\text{or}\,\,\, f|_{02}=f_{226}.
\end{equation}
For configurations restricted to $\{1,2\}$ we obtain
\begin{equation}\label{c3}
f|_{12}=f_{184} \,\,\,\text{or}\,\,\, f|_{12}=f_{226}.
\end{equation}
A quick glance at Table~\ref{allconserving15} convinces us that there is no all-conserving rule
satisfying simultaneously conditions of eq. (\ref{c1}), (\ref{c2}), and (\ref{c3}).
This demonstrates that an all-conserving $f$ with the desired properties does not exist. 

For absolute majority DCP, the proof is identical, as absolute majority and simple majority are the same when
we restrict configurations to two states only. $\square$
 \begin{table}
\begin{center}
   \begin{tabular}{|c|c|c|c|} \hline
Wolfram number $f$ & $f|_{01}$ & $f|_{02}$ & $f|_{12}$ \\ \hline 
  6213370633533  & 204 & 184 & 184 \\
  6768185473053  & 204 & 204 & 184 \\
  6924717700245  & 204 & 184 & 204 \\
  7479532539765  & 204 & 204 & 204 \\
  7486506443925  & 204 & 226 & 204 \\
  7573493966013  & 204 & 204 & 226 \\
  7580467870173  & 204 & 226 & 226 \\
  6914257071453  & 184 & 184 & 204 \\
  7469071910973  & 184 & 204 & 204 \\
  7563033337221  & 184 & 204 & 226 \\
  6769347793221  & 226 & 204 & 184 \\
  7480694859933  & 226 & 204 & 204 \\
  7487668764093  & 226 & 226 & 204 \\
  7625403764901  & 240 & 240 & 240 \\
  6159136430181  & 170 & 170 & 170 \\ \hline
  \end{tabular}
\end{center}
\caption{List of the Wolfram codes of the 15 all-conserving ternary rules (first column). Wolfram numbers of their binary projections are shown in columns
$f|_{01}$, $f|_{02}$, and $f|_{12}$.} \label{allconserving15}
\end{table}

\section{Interval-wise density classification}
In addition to the variants of the DCP described in previous section, there exist yet another
possible way to generalize the density classification problem which includes the ``classical'' binary DCP as a special case.

Suppose we want to classify finite strings of length $L$ over the alphabet of $M$ symbols ${\cal A}=\{0, 1, \ldots, M-1\}$.
Let 
$$\rho(\mathbf{x})=\frac{\sum_{i=0}^{L-1} x_i}{\sum_{i=0}^{L-1} \mathrm{max}\, \cal {A}}$$ 
be called \emph{density} of
configuration $\mathbf{x}=(x_0,x_2, \ldots. x_{L-1})$, where $\mathrm{max}\, \cal {A}$ is the largest element of
the alphabet ${\cal A}$,  so that  $\mathrm{max}\, {\cal A}= M-1$. This definition guarantees that  $\rho \in [0,1]$,
and obviously
$$\rho(\mathbf{x})=\frac{1}{L(M-1)} \sum_{i=0}^{L-1} x_i.$$ 
Let $p_1,p_2, \ldots, p_{M-1}$ be real numbers satisfying $0<p_1<p_2< \ldots <p_{M-1} <1$. The pair of rules
with global functions $F$ and $G$ solves interval-wise density classification problem if there exist integers $n,m$ (possibly depending on $L$)
such that  for every configuration  $\mathbf{x}=(x_0,x_2, \ldots. x_{L-1})$ 
we have
\begin{align*}
 G^mF^n(\mathbf{x})&=0^L \text{\,\,if\,\,} \rho(\mathbf{x}) \in [0, p_1),\\
 G^mF^n(\mathbf{x})&=1^L \text{\,\,if\,\,} \rho(\mathbf{x}) \in (p_1, p_2),\\
G^mF^n(\mathbf{x})&=2^L \text{\,\,if\,\,} \rho(\mathbf{x}) \in (p_2, p_3),\\
 & \cdots\\
 G^mF^n(\mathbf{x})&=(M-1)^L \text{\,\,if\,\,} \rho(\mathbf{x}) \in (p_{M-1}, 1].
\end{align*}
Obviously, when $M=2$ and $p_1=1/2$, the above reduces to the ``classical'' binary density classification problem,
with known solution by the pair of ECA 184 and 232. Note that we intentionally made the intervals open at points 
$p_1, p_2, \ldots p_{M-1}$, to be compatible with the standard DCP, where the classification
of strings with equal number of zeros and ones is not required.

When $p_i=1/M$ for $i=1, 2, \ldots M-1$, we call the problem \emph{symmetric interval-wise} density classification
problem. For ternary rules ($M=3$), does a two-rule solution of the symmetric interval-wise DCP exist?
Again, based on our extensive heuristic search, we suspect that the answer is no, but
we were able to prove the non-existence only for rules reducible to two states.

Suppose that such solution indeed existed, consisting of rules with global functions $F$ and $G$, both reducible to
two states. This would mean that for some $m$ and $n$ we have
\begin{align*}
 G^mF^n(\mathbf{x})&=0^L \text{\,\,if\,\,} \rho(\mathbf{x}) \in [0, 1/3),\\
 G^mF^n(\mathbf{x})&=1^L \text{\,\,if\,\,} \rho(\mathbf{x}) \in (1/3, 2/3),\\
 G^mF^n(\mathbf{x})&=2^L \text{\,\,if\,\,} \rho(\mathbf{x}) \in (2/3, 1].
\end{align*}
Now let us suppose that $\mathbf{x}$ is binary, consisting only of 0's and 1's. In such a case,
the above would reduce to
\begin{align*}
 G^mF^n(\mathbf{x})&=0^L \text{\,\,if\,\,} 0 \leq \rho(\mathbf{x}) < 1/3,\\
 G^mF^n(\mathbf{x})&=1^L \text{\,\,if\,\,} 1/3 < \rho(\mathbf{x}) < 2/3,
\end{align*}
or, using the definition of $\rho(\mathbf{x})=\frac{1}{2L}\sum x_i$,
\begin{align*}
 G^mF^n(\mathbf{x})&=0^L \text{\,\,if\,\,} 0 \leq \frac{1}{L}\sum_{i=0}^{L-1} x_i < 2/3,\\
 G^mF^n(\mathbf{x})&=1^L \text{\,\,if\,\,} 1/3 < \frac{1}{L}\sum_{i=0}^{L-1} x_i < 4/3.
\end{align*}
This would imply existence of a pair of elementary rules having the property
that $G^mF^n(\mathbf{x})$ consists of all zeros  if zeros occupy $2/3$ of all sites of the initial configuration
and all ones otherwise. No such pair of elementary rules exists, and it can be verified numerically by checking all possible cases,
as there are only 256 elementary rules, yielding $256 \cdot 256=65536$ possible pairs. This proves the following.
\begin{proposition}
 There is no pair of nearest-neighbour ternary rules reducible to two states which would solve the symmetric interval-wise density classification problem.
\end{proposition}
\section{Non-symmetric interval-wise classification}
If there is no solution of the symmetric interval-wise problem, can we at least perform non-symmetric classification
with two ternary rules?

We performed an intensive heuristic search for such rules, and after some tinkering with rule tables, by trial and error we found the following interesting pair of ternary rules. 

\begin{conjecture}\label{nonsymintervalclass}
Let
 $F$ be the ternary nearest-neighbour rule with Wolfram number 6478767664173, and
 $G$  be the rule with Wolfram number  7580606234490. 
 For any finite ternary string $\mathbf{x}$  of length  $L$, containing at least one zero, let $\displaystyle \rho=\frac{1}{2L}\sum_{i=0}^{L-1} x_i$. Then
\begin{align*}
 G^LF^L(\mathbf{x})&=0^L \text{\,\,if\,\,} \rho(\mathbf{x}) \in [0, 2/3),\\
 G^LF^L(\mathbf{x})&=1^L \text{\,\,if\,\,} \rho(\mathbf{x}) \in (2/3, 3/4),\\
 G^LF^L(\mathbf{x})&=2^L \text{\,\,if\,\,} \rho(\mathbf{x}) \in (3/4, 1).
\end{align*}
\end{conjecture}
This means that the pair of rules (6478767664173, 7580606234490) ``almost'' solves the interval-wise DCP with $p_1=2/3$ and
$p_2=3/4$. We use the word ``almost'' because it only works for configurations which contain at least one zero.
Configurations which do not satisfy this property can be misclassified - for example, $1^L$ has density $1/2$, thus should
produce  $0^L$ in the end, yet $1^L$ is a fixed point of both rules 6478767664173 and 7580606234490\footnote{We wish
to thank anonymous referee for pointing out this fact. }.

We performed extensive numerical experiments to verify the above conjecture, and it appears to be valid.
Below we provide a sketch of a possible proof of this result. 
\begin{figure} 
\begin{center}
\small
\begin{minipage}[b]{2.5cm}
\begin{align*}
f(0,0,0)&=0\\
f(0,0,1)&=0\\
f(0,1,0)&=1\\
f(0,1,1)&=2\\
f(1,0,0)&=0\\
f(1,0,1)&=0\\
f(1,1,0)&=0\\
f(1,1,1)&=1
\end{align*}
\end{minipage}
\begin{minipage}[b]{2.5cm}
\begin{align*}
f(0,0,0)&=0\\
f(0,0,2)&=1\\
f(0,2,0)&=1\\
f(0,2,2)&=2\\
f(2,0,0)&=0\\
f(2,0,2)&=1\\
f(2,2,0)&=1\\
f(2,2,2)&=2
\end{align*}
\end{minipage}
\begin{minipage}[b]{2.5cm}
\begin{align*}
f(1,1,1)&= \mathbf{1}\\
f(1,1,2)&= \mathbf{1}\\
f(1,2,1)&= \mathbf{1}\\
f(1,2,2)&= \mathbf{2}\\
f(2,1,1)&= \mathbf{2}\\
f(2,1,2)&= \mathbf{2}\\
f(2,2,1)&= \mathbf{1}\\
f(2,2,2)&= \mathbf{2}
\end{align*}
\end{minipage}
\begin{minipage}[b]{2.5cm}
\begin{align*}
f(0,1,2)&=2\\
f(0,2,1)&=1\\
f(1,0,2)&=1\\
f(1,2,0)&=1\\
f(2,0,1)&=0\\
f(2,1,0)&=1
\end{align*}
\end{minipage}
\end{center}
\caption{Definition of rule 6478767664173. Entries with bold outputs correspond to
restriction to configurations consisting only of 1's and 2's, on which the above rule is equivalent
to rule 184.} \label{rule1}
\end{figure}
\begin{figure}
 \begin{center}  
\small
\begin{minipage}[b]{2.5cm}
\begin{align*}
f(0,0,0)&=0\\
f(0,0,1)&=0\\
f(0,1,0)&=0\\
f(0,1,1)&=0\\
f(1,0,0)&=0\\
f(1,0,1)&=0\\
f(1,1,0)&=1\\
f(1,1,1)&=1
\end{align*}
\end{minipage}
\begin{minipage}[b]{2.5cm}
\begin{align*}
f(0,0,0)&=0\\
f(0,0,2)&=0\\
f(0,2,0)&=0\\
f(0,2,2)&=2\\
f(2,0,0)&=0\\
f(2,0,2)&=2\\
f(2,2,0)&=2\\
f(2,2,2)&=2
\end{align*}
\end{minipage}
\begin{minipage}[b]{2.5cm}
\begin{align*}
f(1,1,1)&=1\\
f(1,1,2)&=1\\
f(1,2,1)&=1\\
f(1,2,2)&=2\\
f(2,1,1)&=1\\
f(2,1,2)&=2\\
f(2,2,1)&=2\\
f(2,2,2)&=2
\end{align*}
\end{minipage}
\begin{minipage}[b]{2.5cm}
\begin{align*}
f(0,1,2)&=0\\
f(0,2,1)&=0\\
f(1,0,2)&=0\\
f(1,2,0)&=2\\
f(2,0,1)&=1\\
f(2,1,0)&=1
\end{align*}
\end{minipage}
 \end{center}
\caption{Definition of rule 7580606234490. Its binary projections are ECA 192, 232, and 232,
and this can be verified by inspecting the first three columns.} \label{rule2}
\end{figure}
Figures~\ref{rule1} and \ref{rule2} show definitions of rules $F$ and $G$. The first of them (rule 6478767664173)
is number-conserving, and its binary projection $f|_{12}$ (defined as in eq. \ref{binaryproj}) is ECA 184.
This rule plays an analogous role as rule 184 in the two-rule solution of the binary DCP, namely,
it prepares the configuration for further processing without changing its density $\rho(\mathbf{x})$.
After sufficiently many iterations (which we simply take to be $L$), this rule eliminates
certain symbols and substrings, as shown in Table~\ref{eliminationtable}.

 One can see that, for example, when $\rho(\mathbf{x})<2/3$,
after $L$ iterations of rule $F$, the configuration may contain substrings 00 and 11 as well as
symbols 0, 1, and 2, while substrings 22 are always absent. When $\rho(\mathbf{x}) \in (2/3, 3/4)$, the
configuration may contain substrings 11 and symbols 1 and 2, while substrings 00 and 22 as well as
0's are absent. And, finally, when $\rho(\mathbf{x})>3/4$, substrings 22 may be present, 00 and 11 are absent,
while symbols 1 and 2 may be present and 0's are absent.
\begin{table}
\begin{center}
\begin{tabular}{l|l|l|l}
substring & $\rho(\mathbf{x})<2/3$ & $\rho(\mathbf{x}) \in (2/3, 3/4)$ & $\rho(\mathbf{x})>3/4$  \\ \hline
00 & Yes & No & No\\
11  & Yes & Yes & No\\
22 & No & No & Yes\\
0 & Yes & No & No\\
1 & Yes & Yes & Yes\\
2 & Yes & Yes & Yes
 \end{tabular}
 \end{center}
\caption{Presence of selected substrings in the final configuration after iterating rule 6478767664173
$L$ times.} \label{eliminationtable}
\end{table}

The second rule $G$, with Wolfram number 6478767664173, plays a role similar to rule 232 in the two-rule solution of
the binary DCP. Its  three binary projections  $f|_{01}$ ,  $f|_{02}$ and $f|_{12}$  exist, and their Wolfram numbers are, respectively,
192, 232, and 232. This rule grows clusters of 0's if they are present, and if not, it just behaves like rule 232,
that is, grows clusters of 1's in the absence of pairs 22, and grows clusters of 2's in the absence of 
pairs 11.
 A quick look at Table~\ref{eliminationtable} reveals that these three cases will occur after we iterate
rule $F$ on initial configurations with densities, respectively, $\rho(\mathbf{x})<2/3$, $\rho(\mathbf{x}) \in (2/3, 3/4)$, and
 $\rho(\mathbf{x})>3/4$.  Note that presence of 0's is required to grow clusters of 0's, thus the need for additional condition
imposed in the initial configuration (it mus contain at least one zero).

\begin{figure}
\includegraphics[width=4.1cm]{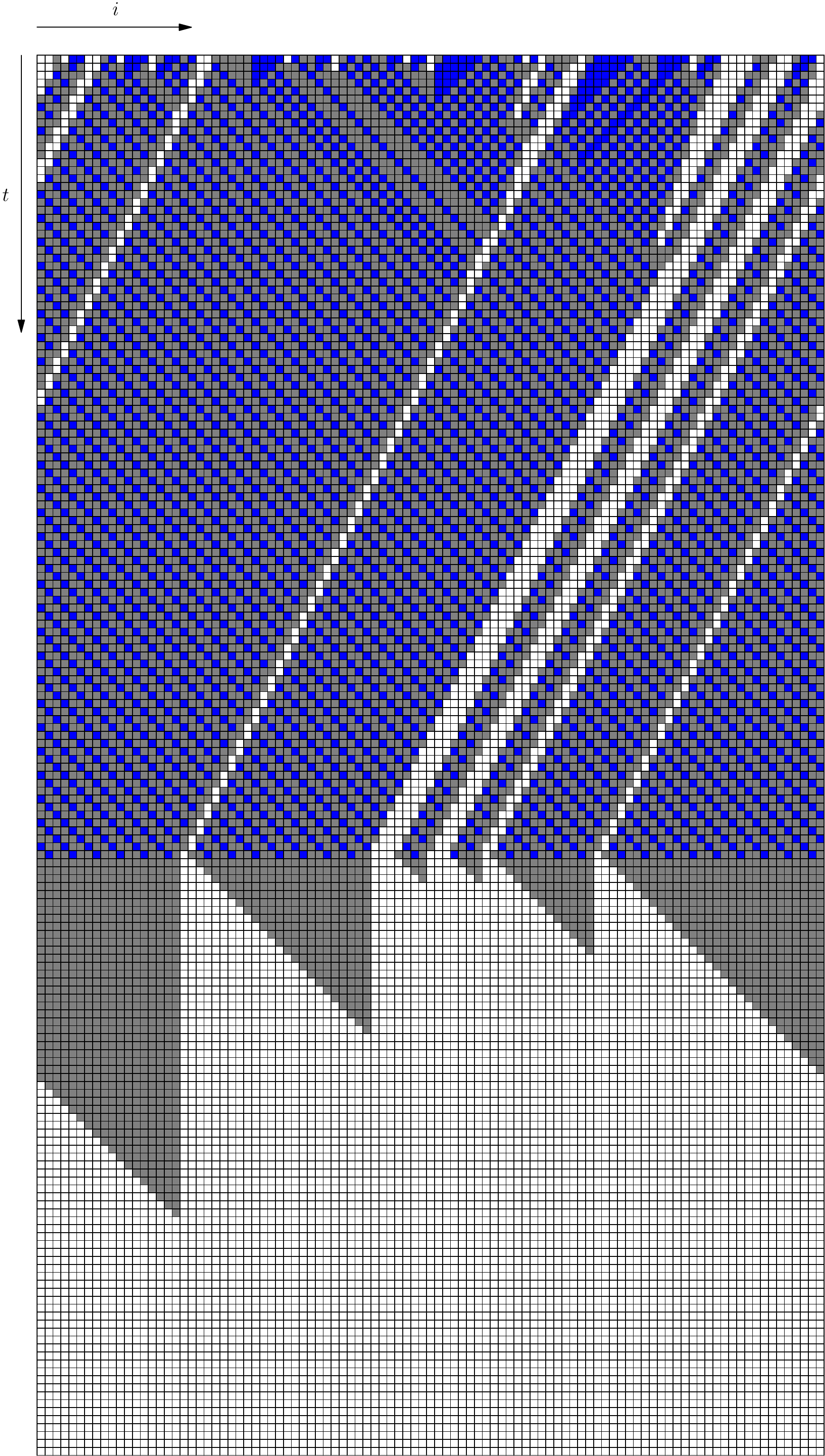}
\includegraphics[width=4.1cm]{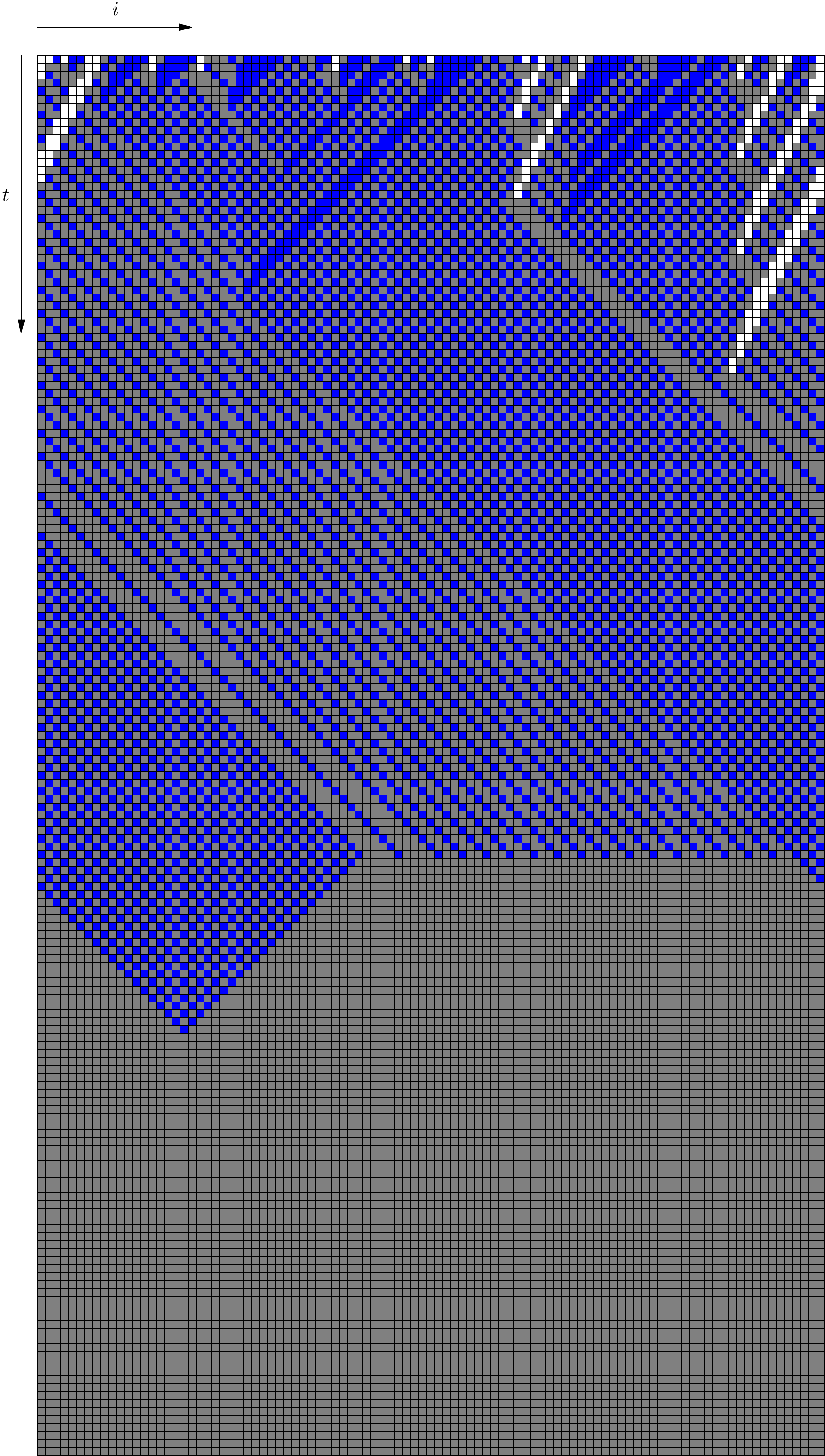}
\includegraphics[width=4.1cm]{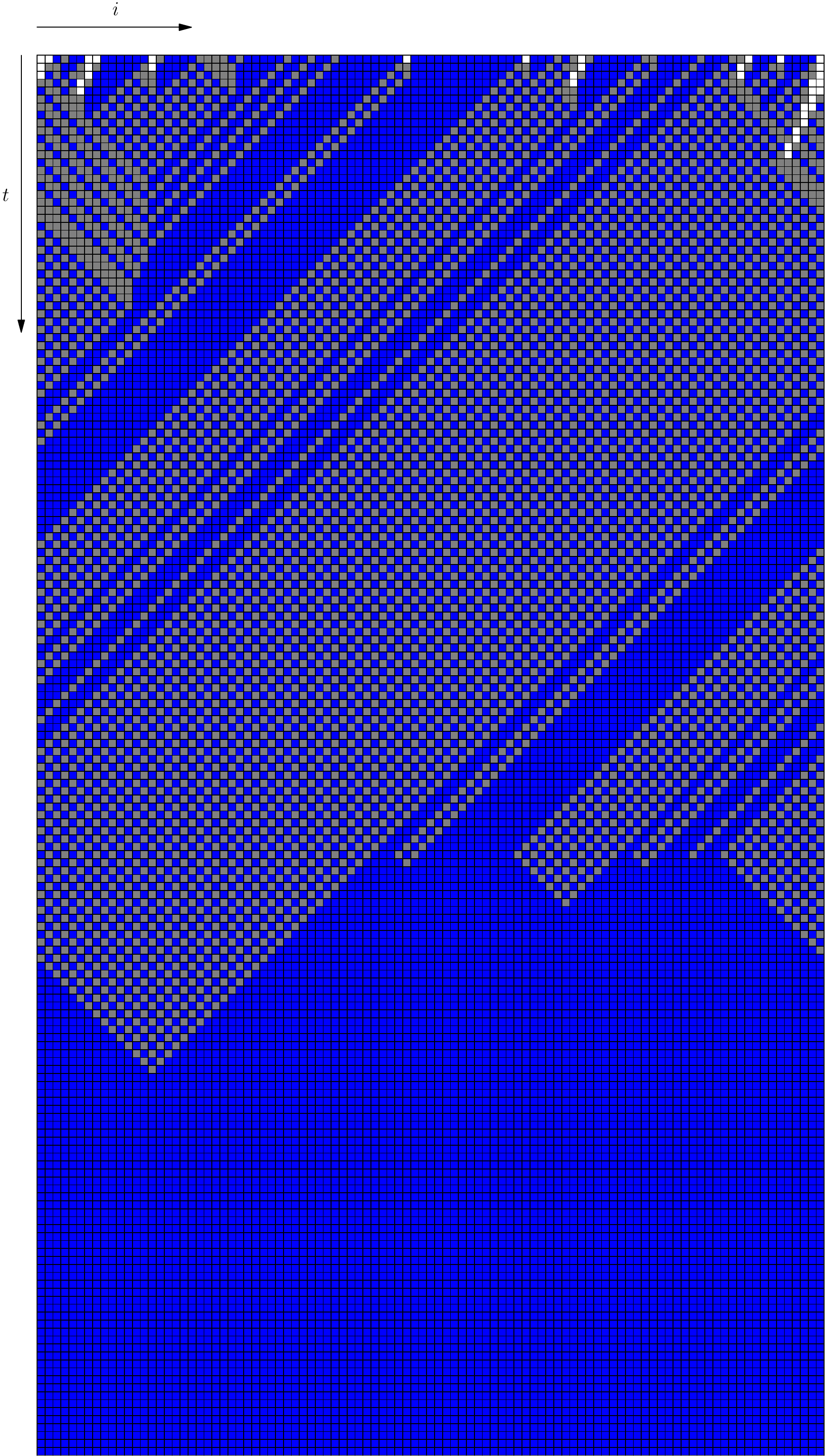}\\
$\rho(\mathbf{x})=0.6$ \hspace{2.5cm} $\rho(\mathbf{x})=0.7$\hspace{2.5cm}$\rho(\mathbf{x})=0.8$
\caption{Spatiotemporal patterns illustrating a two-rule solution of non-symmetric interval-wise DCP by rules
6478767664173 and 7580606234490. White color represents 0's, gray 1's, and blue 2's.}  \label{patterns}
\end{figure}

 The final effect, therefore,  of  iterations
of rule $G$ starting with $F^L(\mathbf{x})$ will be all zeros when $\rho(\mathbf{x})<2/3$, all ones when $\rho(\mathbf{x}) \in (2/3, 3/4)$,
and all 2's when $\rho(\mathbf{x})>3/4$, exactly as claimed in Conjecture~\ref{nonsymintervalclass}.
Examples of  three cases of density classification by the aforementioned rules are shown in Figure~\ref{patterns}.

Obviously the above is only a sketch of a proof, and it needs further elaboration. Our statement about rules
6478767664173 and 7580606234490, therefore, must remain a conjecture for now.

\section{Conclusions and future work}
We have demonstrated that except the trivial case of 0-majority, there exist no two-rule solution of 
various density classification problems (simple majority, absolute majority, symmetric interval-wise) in the domain
of ternary nearest-neighbour rules  which would be analogous to the known solution of DCP by the pair of ECA 184 and 232.
By ``analogous'' we mean a solution consisting of two rules reducible to two states in which the first rule serves as a pre-processor preserving relationship between
densities, thus is all-conserving for simple/absolute majority problem, or number-conserving for interval-wise problem. 

This naturally brings up a question if two-rule solutions exist if one relaxes the restriction of the first rule possessing additive invariant or the restriction of rules being reducible. While such a possibility cannot be excluded, we seriously doubt it. 
However, it seems quite possible that extending the neighbourhood size to two nearest neighbours may  help to produce a two-rule solution. We plan to investigate this
possibility in the near future.

\begin{scriptsize}\noindent\textbf{Acknowledgement:} H.F. acknowledges financial support from the Natural Sciences and
Engineering Research Council of Canada (NSERC) in the form of Discovery Grant.  We thank anonymous referees for
comments which helped to improve the paper. 
\end{scriptsize}

\providecommand{\href}[2]{#2}\begingroup\raggedright\endgroup
\end{document}